\documentclass[aps,pre,twocolumn,showpacs,superscriptaddress]{revtex4}
\usepackage{epsfig}
\usepackage{times}

\begin{document}

\title{Evolution of extortion in structured populations}

\author{Attila Szolnoki}
\affiliation{Institute of Technical Physics and Materials Science, Research Centre for Natural Sciences, Hungarian Academy of Sciences, P.O. Box 49, H-1525 Budapest, Hungary}

\author{Matja{\v z} Perc}
\affiliation{Faculty of Natural Sciences and Mathematics, University of Maribor, Koro{\v s}ka cesta 160, SI-2000 Maribor, Slovenia}

\begin{abstract}
Extortion strategies can dominate any opponent in an iterated prisoner's dilemma game. But if players are able to adopt the strategies performing better, extortion becomes widespread and evolutionary unstable. It may sometimes act as a catalyst for the evolution of cooperation, and it can also emerge in interactions between two populations, yet it is not the evolutionary stable outcome. Here we revisit these results in the realm of spatial games. We find that pairwise imitation and birth-death dynamics return known evolutionary outcomes. Myopic best response strategy updating, on the other hand, reveals new counterintuitive solutions. Defectors and extortioners coarsen spontaneously, which allows cooperators to prevail even at prohibitively high temptations to defect. Here extortion strategies play the role of a Trojan horse. They may emerge among defectors by chance, and once they do, cooperators become viable as well. These results are independent of the interaction topology, and they highlight the importance of coarsening, checkerboard ordering, and best response updating in evolutionary games.
\end{abstract}

\maketitle

Cooperation is an evolutionary riddle, as it defies the basic principles of natural selection. If during the course of evolution only the fittest survive, why should one sacrifice individual fitness for the benefit of unrelated others? Widespread cooperation in nature is indeed one of the most important challenge to Darwin's theory of evolution and natural selection. Understanding the evolution of cooperation means understanding also the main evolutionary transitions that led from single-cell organisms to complex animal and human societies \cite{maynard_95}, and it is therefore little surprising that the subject consistently attracts attention across large contingents of social and natural sciences.

Evolutionary game theory \cite{maynard_82, weibull_95, hofbauer_98, nowak_06} is frequently employed as the theoretical framework within which we try to understand and explain the evolution of cooperation. Prominent mechanism that promote cooperative behavior are kin selection \cite{hamilton_wd_jtb64a}, network reciprocity \cite{nowak_n92b}, direct and indirect reciprocity \cite{trivers_qrb71, axelrod_s81}, as well as group \cite{wilson_ds_an77} and multilevel selection \cite{traulsen_pnas06, szolnoki_njp09}, all of which were comprehensively reviewed as the five rules for the evolution of cooperation in \cite{nowak_s06}. There are also a number of related reviews devoted to evolutionary games that survey succinctly recent advances made along this vibrant avenue of research \cite{szabo_pr07, sigmund_tee07, roca_plr09, perc_bs10, perc_jrsi13, rand_tcs13}.

The iterated prisoner's dilemma game is probably the most often used workhorse for studying the evolution of cooperation among selfish individuals \cite{fudenberg_e86, nowak_n93, szabo_pre98, santos_prl05, imhof_pnas05, tanimoto_pre07, gomez-gardenes_prl07, poncela_njp07, fu_pre08b, poncela_epl09, lee_s_prl11, antonioni_pone11, tanimoto_pre12}. The famous tournaments organized by Robert Axelrod \cite{axelrod_84} have revealed that tit-for-tat is the most successful strategy. Similar to retaliation or reciprocity, the virtue of the tit-for-tat strategy is to follow the opponents previous action, although initially to always cooperate. Only few concepts were thus far able to challenge the success of this simple but effective strategy \cite{nowak_n93}. Press and Dyson have recently introduced so-called zero-determinant strategies \cite{press_pnas12}. These strategies impose a linear relation between one's own payoff and the payoff of the other player. Extortion strategies are a subset of zero-determinant strategies, which furthermore ensure that an increase in one's own payoff exceeds the increase in the other player's payoff by a fixed percentage. Extortion strategies are therefore able to dominate any evolutionary opponent, including tit-for-tat and in fact all other strategies \cite{stewart_pnas12}. But in the realm of evolutionary games, where players are able to imitate strategies that are performing better, extortion quickly becomes widespread and in fact evolutionary unstable \cite{adami_ncom13}. If everybody extorts, it is better to cooperate. The outlook for extortioners, however, is not quite so bleak, especially if the two players engaged in the game belong to distinct populations, or if the population size is very small \cite{hilbe_pnas13}. It is also possible to devise generous zero-determinant strategies, which support each other and are therefore evolutionary stable \cite{stewart_pnas13}.

Here we continue to explore the evolutionary viability of extortion, but instead of well-mixed populations, we focus on games in structured populations \cite{szabo_pr07}. By doing so, we take into account the fact that the interactions among players are frequently not random and best described by a well-mixed model, but rather that they are limited to a set of other players in the population and as such are best described by a network. We consider a $L \times L$ square lattice with periodic boundary conditions as the simplest of networks to fulfill this condition, as well as the scale-free network with the same average degree, which is likely a more apt model for realistic social and technological networks \cite{barabasi_s99}. As we will show, however, the main results remain unaffected by the topological differences of the interaction networks.

In terms of game parametrization, we follow closely the work of Hilbe et al. \cite{hilbe_pnas13}, where extortion was studied in the realm of the donation game. The latter is a special case of the iterated prisoner's dilemma game, which however retains all the original properties of the social dilemma. The competing strategies are cooperation $C$, defection $D$, and extortion $E_\chi$. The payoff matrix is
\begin{equation}
\begin{tabular}{r|c c c}
 & $E_\chi$ & $C$ & $D$ \\
\hline
$E_\chi$ & 0 & $\frac{(b^2-c^2)\chi}{b\chi + c}$ & 0 \\
$C$ & $\frac{b^2-c^2}{b\chi+c}$ & $b-c$ & $-c$ \\
$D$ & 0 & $b$ & 0 \\
\end{tabular}
\end{equation}
where $b$ is the benefit to the other player provided by each cooperator at the cost $c$, and $\chi$ determines the surplus of the extortioner in relation to the surplus of the other player. Moreover, we use $b-c=1$, thus having $b>1$ and $\chi>1$ as the two main parameters. The former determines the strength of the social dilemma, while the latter determines just how strongly strategy $E_\chi$ exploits cooperators. It is worth noting that we are focusing on the simplest three-strategy model and thus do not consider strategies such as tit-for-tat \cite{szolnoki_pre09b} or win-stay-lose-learn \cite{liu_yk_pone12} or generous zero-determinant strategies \cite{stewart_pnas13}. Partly this is because the evolutionary success of some of these strategies has already been studied thoroughly in structured populations, but also because we wish to keep the analysis as conclusive and as clear as possible with regards to the evolutionary prospects of extortion.

\begin{figure}
\centerline{\epsfig{file=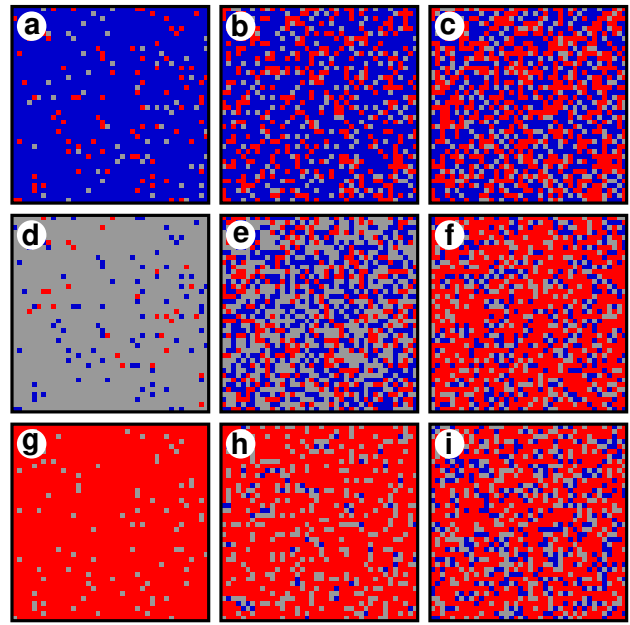,width=8.5cm}}
\caption{Spatial distributions of cooperators (blue), defectors (red) and extortioners (gray), as obtained from three different initial conditions on the square lattice by means of myopic best response strategy updating. (a-c) Evolution starts from a full $C$ phase. Extortioners and defectors can emerge because they are both superior in the sea of cooperators. But defectors are more probable due to their relatively higher payoffs. (d-f) Evolution starts from a full $E_\chi$ phase. Defectors emerge by chance since they are neutral with extortioners. Cooperators also emerge because they outperform extortioners if the latter are in majority. (g-i) Evolution starts from a full $D$ phase. Here $E_\chi$ emerge by chance since they are neutral with defectors. As soon as extortioners segregate and form small compact domains, cooperators become viable too. The pure $D$ phase thus erodes to give rise to a stable mixed $C+D+E_\chi$ phase that sets in regardless of the initial conditions (c,f,i). Parameter values in all three cases are $b=1.5$, $\chi=1.5$, $K=0.05$, and $L=40$. Initial homogeneous states are not shown.}
\label{snapshot}
\end{figure}

Unless stated differently, for example to illustrate specific invasion processes as in Fig.~\ref{snapshot}, we use random initial conditions such that all three strategies are uniformly distributed across the network. We carry out Monte Carlo simulations comprising the following elementary steps. First, a randomly selected player $x$ with strategy $s_x$ acquires its payoff $p_x$ by playing the game with its $k$ neighbors, as specified by the underlying interaction network. Next, player $x$ changes its strategy $s_x$ to $s_x^{\prime}$ with the probability
\begin{equation}
q(s_x^{\prime} \to s_x) =\frac{1}{1+\exp[(p_x-p_x^{\prime})/K]}\ \,
\label{myop}
\end{equation}
where $p_x^{\prime}$ is the payoff of the same player if adopting strategy $s_x^{\prime}$ within the same neighborhood, and $K$ is the uncertainty related to the strategy adoption process \cite{szabo_pr07}. The strategy $s_x^{\prime}$ should of course be different from $s_x$, and it is drawn randomly from the remaining two strategies. Such strategy updating is known as the myopic best response rule \cite{matsui_jet92}. We also consider the more traditional strategy imitation, where player $x$ imitates the strategy of a randomly selected neighbor $y$, only that $p_x^{\prime}$ in Eq.~\ref{myop} is replaced by $p_y$ \cite{szabo_pre98}, as well as death-birth updating as described for example in \cite{ohtsuki_jtb06}. Regardless of the applied strategy updating rule, we let the system evolve towards the stationary state where the average frequency of strategies becomes time independent.

The results obtained via strategy imitation and birth-death updating are quickly explained, and they are in fact qualitatively in agreement with the results obtained on well-mixed populations in that extortion strategies face a rather gloomy evolutionary outlook \cite{adami_ncom13, hilbe_pnas13}. In particular, if $b$ is low enough for cooperators to survive in the presence of defectors (which would be due to network reciprocity \cite{nowak_n92b}), then $E_\chi$ always die out regardless of $\chi$. On the other hand, if $b$ is too high for cooperators to survive, the remaining defectors and extortioners become neutral. But since $D$ are in general more successful in invading cooperators than $E_\chi$, the majority of players at the time of cooperation extinction will have strategy $D$. The absorbing $D$ phase is therefore a much more likely final evolutionary outcome of logarithmically slow coarsening \cite{dornic_prl01} then the absorbing $E_\chi$ phase. Overall, extortion is unable to capitalize on structured interactions if the strategy updating is governed by imitation or a birth-death rule.

\begin{figure}
\centerline{\epsfig{file=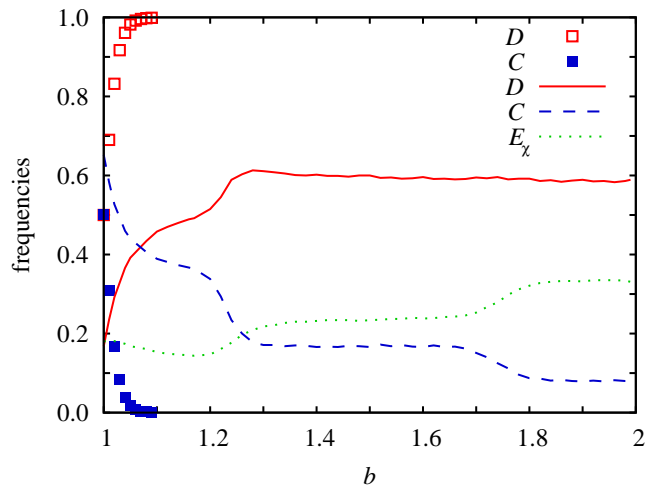,width=8.5cm}}
\caption{Myopic best response strategy updating preserves cooperation across the whole interval of the temptation to defect $b$ if extortioners are part of the game (lines). If solely cooperators and defectors compete, the latter rise to complete dominance already at $b=1.09$ (symbols). Extortion thus catalyzes cooperation in structured populations. Here we have used $\chi=1.5$ to define the extortion strategy $E_\chi$. Figure~\ref{chi-b} shows results for the whole $b-\chi$ plane and for two different interaction networks.}
\label{chi1_5}
\end{figure}

Myopic strategy updating, on the other hand, reveals very different behavior as it allows players to adopt strategies that are not necessarily present in their interaction neighborhood. To begin with, it is worth emphasizing that such strategy updating is not equivalent with mutation because each individual update is still driven by the payoffs (see Eq.~\ref{myop}). The difference compared  to mutation-driven evolution can be illustrated nicely with the traditional two-strategy version of the prisoner's dilemma game, where cooperators always die out above a critical temptation to defect. As depicted in Fig.~\ref{chi1_5} (symbols), the frequency of cooperators goes to zero at $b=1.09$. We also note that in this paper, to avoid the potentially disturbing impact of noise, we have used $K=0.05$ in Eq.~\ref{myop}, which practically prevents a strategy change if the new strategy does not yield a higher payoff.

\begin{figure}
\centerline{\epsfig{file=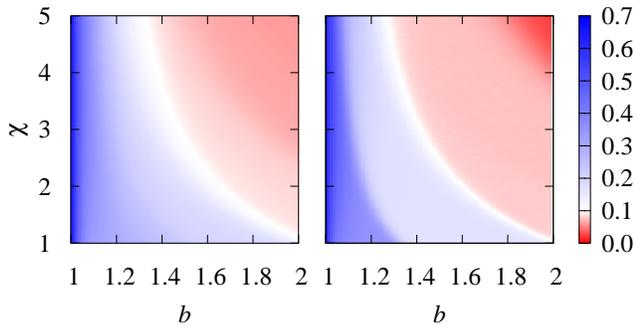,width=8.5cm}}
\caption{Color-encoded stationary frequency of cooperators on the whole $b-\chi$ parameter plane as obtained on the scale-free network (left) and the square lattice (right) by means of myopic best response strategy updating. Cooperators are able to survive across the whole interval of the temptation to defect $b$ as long as $\chi$ is not too large. This outcome is independent of the interaction topology since results obtained on the square lattice and on the scale-free network are to a large extent the same.}
\label{chi-b}
\end{figure}

Unexpectedly, if all three strategies compete, extortion provides an evolutionary escape hatch for cooperators to survive even at the most prohibitive conditions ($b=2$), as illustrated in Fig.~\ref{chi1_5} (lines). This result is counterintuitive because the introduction of extortioners increases the number of those who exploit cooperators. Although extortion acts more subtly than defection, it is still difficult to imagine how it can promote cooperation. Moreover, extortion itself becomes evolutionary stable, and at sufficiently large $b$ even outperforms cooperation. Results presented in Fig.~\ref{chi-b} add further support to these claims, evidencing that extortion indeed always supports some level of cooperation, as long as $\chi$ is within reasonable bounds and $b<2$, and this independently of the topology of the interaction network.

\begin{figure}
\centerline{\epsfig{file=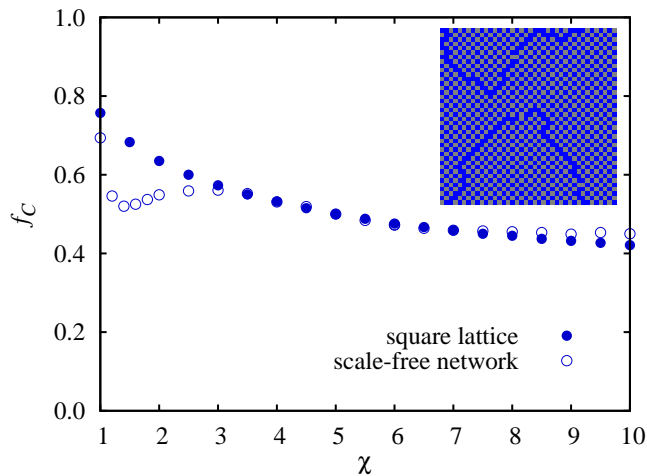,width=8.5cm}}
\caption{In the absence of defectors, cooperators and extortioners are effectively engaged in a snowdrift game. This relation between strategies $C$ and $E_\chi$ results in checkerboard ordering, where players do not have to interact with others of the same kind (see inset). The stationary fraction of cooperators $f_C$ therefore remains high regardless of $\chi$, and regardless of the topology of the interaction network (main panel). Importantly, the separation emerges spontaneously due to the snowdrift relation and myopic best response strategy updating. Parameter values are $b=1.5$ (main panel and inset) and $\chi=2$ (inset).}
\label{c-ext}
\end{figure}

To explain these results, we monitor the evolution of the distribution of strategies from three different homogeneous states. The top row of Fig.~\ref{snapshot} starts from a full $C$ phase. Expectedly, defection emerges fast as does extortion, since both strategies offer higher payoffs than cooperation in the neighborhood of other cooperators. The middle row of Fig.~\ref{snapshot} depicts the evolution from a full $E_\chi$ phase. Here defectors emerge by chance as they are neutral with extortioners, but cooperators emerge because their payoff is higher in the sea of extortioners. Interestingly, if strategy $D$ would not be an alterative, a checkerboard configuration would emerge spontaneously, where $C$ and $E_\chi$ are able to support each other due to their snowdrift-like relation (see the inset of Fig.~\ref{c-ext}). But the availability of strategy $D$ destroy this ordering, instead giving rise to a mixed $C+D+E_\chi$ phase. The most interesting, however, is the erosion of the full $D$ phase depicted in the bottom row of Fig.~\ref{snapshot}. Here initially only extortioners emerge by chance since cooperators are obviously not competitive. Yet the emergence of $E_\chi$ allows cooperators to appear as well. More precisely the coarsening of $D$ and $E_\chi$ players will result in small homogeneous $E_\chi$ clusters, which creates the chance for cooperators to appear. In this way, extortion thus plays the role of a Trojan Horse and helps cooperators to conquer defector-dominant areas. Nevertheless, the spreading of cooperation, which utilizes the neutral drift of $E_\chi$, will be controlled by defectors who can strike back since their presence in place of an extortioner may yield a higher payoff in a predominantly cooperative neighborhood. Temporarily this is certainly the case, but soon thereafter other players within the neighborhood will start changing their strategies too, eventually arriving at the pure $D$ (at least locally) zero-productivity state. From this point onwards extortioners will start reappearing through neutral drift and essentially restart the whole cycle of dominance again. The stationary mixed $C+D+E_\chi$ phase, as is specific for the applied value of $b$ and $\chi$, ultimately sets in as a consequence of the described elementary evolutionary invasions.

Although extortion can be as counterproductive as defection, it is still less destructive. For a cooperator it never pays sticking with the strategy if surrounded by defectors, but it may be the best option among extortioners. Evidently, cooperators are happiest among other cooperators, but in the presence of extortioners they can still attain a positive payoff, and this is much better than nothing or a negative value in the presence of defectors. Accordingly, in a homogeneous population of extortioners it is better to deviate by cooperating \cite{hilbe_pnas13}. Although in structured populations this change always happens locally, it can also be observed globally in a two-strategy game entailing only $C$ and $E_\chi$ strategies. As illustrated in the inset of Fig.~\ref{c-ext}, the snowdrift relation gives rise to a checkerboard ordering, where extortioners do not have to interact with players of their own kind. We note that similar ordering was already observed in traditional $C-D$ spatial games under myopic updating \cite{sysiaho_epjb05}, yet it cannot be observed under imitation dynamics, unless the imitation does not apply to strategy but rather to a different determinant of behavior, such as emotions \cite{szolnoki_epl11}. Importantly, the role-separating coexistence of $C$ and $E_\chi$ players is not restricted to low $\chi$ values, and it is also independent of the interaction topology, as evidenced in the main panel of Fig.~\ref{c-ext}.

To sum up, extortion is evolutionary stable in structured populations if the strategy updating is governed by a myopic best response rule. Counterintuitively, the stability of extortioners helps cooperators to survive even under the most testing conditions, whereby the neutral drift of $E_\chi$ players serves as the entry point, akin to a Trojan horse, for cooperation to grab a hold among defectors. The mutually rewarding checkerboard-like coexistence of cooperators and extortioners can always be temporarily disturbed by defectors since they may earn more in the same neighborhood. But this does not last long since the neighborhood is soon to follow, thus yielding a configuration with zero productivity. The neutral drift, i.e., coarsening in the spatial system, then reintroduce extortioners, and the whole cycle starts anew. The exploration of extortion by means of myopic updating thus offers an unlikely evolutionary niche for the evolution of cooperation, and so it highlights the potential importance of best response updating that is arguably an integral part of human behavior \cite{traulsen_pnas10}.

\begin{acknowledgments}
This research was supported by the Hungarian National Research Fund (Grant K-101490) and the Slovenian Research Agency (Grant J1-4055).
\end{acknowledgments}

\end{document}